\documentclass[journal,draftclsnofoot,onecolumn]{IEEEtran}

\usepackage{setspace}
\usepackage{amsmath}
\usepackage{amssymb}
\usepackage{mathrsfs}
\usepackage{amsthm}
\usepackage{multirow}
\usepackage{color}
\usepackage{longtable}
\usepackage{array}
\usepackage{url}
\usepackage{comment}
\usepackage{enumerate}
\usepackage{eucal}
\usepackage{graphics}

\usepackage{algorithm}
\usepackage{algpseudocode}
\usepackage[noadjust]{cite}
\usepackage[table]{xcolor}

\usepackage{stmaryrd}

\usepackage{mathtools}
\usepackage{xparse}
\DeclarePairedDelimiterX{\set}[1]{\{}{\}}{\setargs{#1}}
\NewDocumentCommand{\setargs}{>{\SplitArgument{1}{;}}m}
{\setargsaux#1}
\NewDocumentCommand{\setargsaux}{mm}
{\IfNoValueTF{#2}{#1} {#1\,\delimsize|\,\mathopen{}#2}}%{#1\:;\:#2}

\DeclarePairedDelimiter\abs{\lvert}{\rvert}

\DeclarePairedDelimiter\parenv{\lparen}{\rparen}
\DeclarePairedDelimiter\sparenv{\lbrack}{\rbrack}

\newcommand{\cA}{\mathcal{A}}
\newcommand{\cB}{\mathcal{B}}
\newcommand{\cC}{\mathcal{C}}

\newcommand{\cE}{\mathcal{E}}
\newcommand{\cF}{\mathcal{F}}

\newcommand{\cS}{\mathcal{S}}

\renewcommand{\leq}{\leqslant}

\renewcommand{\geq}{\geqslant}

\newcommand{\ppmod}[1]{~({\rm mod~}#1)}

\newcommand{\ord}{\log}
\theoremstyle{plain}
\newtheorem{theorem}{Theorem}

\newtheorem{lemma}[theorem]{Lemma}
\newtheorem{proposition}[theorem]{Proposition}

\theoremstyle{definition}
\newtheorem{definition}[theorem]{Definition}

\newcommand{\F}{\mathbb{F}}

\newcommand{\C}{\mathbb{C}}
\newcommand{\Z}{\mathbb{Z}}
\newcommand{\N}{\mathbb{N}}

\newcommand{\ve}{\mathbf{e}}
\newcommand{\vm}{\mathbf{m}}
\newcommand{\vr}{\mathbf{r}}
\newcommand{\vs}{\mathbf{s}}
\newcommand{\vv}{\mathbf{v}}

\newcommand{\vx}{\mathbf{x}}

\newcommand{\va}{\mathbf{a}}
\newcommand{\vb}{\mathbf{b}}
\newcommand{\vc}{\mathbf{c}}

\newcommand{\vh}{\mathbf{h}}

\newcommand{\One}{{\mathbf{1}}}

\newcommand{\kp}{k_+}
\newcommand{\km}{k_-}

\newcommand{\BURB}{{\mathcal E}(n,b,\kp,\km)}
\newcommand{\CBURB}{{\mathcal E}^{\circ}(n,b,\kp,\km)}
\newcommand{\eqdef}{\triangleq}

\title{Perfect Codes Correcting a Single Burst of Limited-Magnitude Errors}

\author{
  Hengjia Wei and Moshe Schwartz,~\IEEEmembership{Senior Member,~IEEE}%
  \thanks{This work was supported in part by the Israel Science Foundation (ISF) under Grant 270/18.}%
  \thanks{This work was submitted in part to ISIT 2022.}%
  \thanks{Hengjia Wei is with the School
    of Electrical and Computer Engineering, Ben-Gurion University of the Negev,
    Beer Sheva 8410501, Israel
    (e-mail: hjwei05@gmail.com).}%
  \thanks{Moshe Schwartz is with the School
    of Electrical and Computer Engineering, Ben-Gurion University of the Negev,
    Beer Sheva 8410501, Israel
    (e-mail: schwartz@ee.bgu.ac.il).}%
}

\begin{document}

\maketitle
\begin{abstract}
Motivated by applications to DNA-storage, flash memory, and magnetic recording, we study perfect burst-correcting codes for the limited-magnitude error channel. These codes are lattices that tile the integer grid with the appropriate error ball. We construct two classes of such perfect codes correcting a single burst of length $2$ for $(1,0)$-limited-magnitude errors, both for cyclic and non-cyclic bursts. We also present a generic construction that requires a primitive element in a finite field with specific properties. We then show that in various parameter regimes such primitive elements exist, and hence, infinitely many perfect burst-correcting codes exist.
\end{abstract}


\section{Introduction}
\IEEEPARstart{I}{n} many communication or storage systems, errors tend to occur in close proximity to each other, rather than occurring independently of each other. If the errors are confined to an interval of positions of length $b$, they are referred to as a \emph{burst  of  length $b$}. Note that not all the positions in the interval are necessarily erroneous.   A code that can  correct  any single burst of
length $b$  is called a \emph{$b$-burst-correcting code}.

The design of burst-correcting codes has been researched  in the error models of substitutions, deletions and insertions. Concerning the substitutions,   Abdel-Ghaffar et al. \cite{AbdMcEOdlTil86,Abd88} showed  the existence of optimum cyclic $b$-burst-correcting codes for any fixed $b$, and Etzion \cite{Etz01b} gave a construction for  perfect binary $2$-burst-correcting codes. As for deletions and insertions, it has been shown in \cite{SchWacGabYaa2017} that correcting a single burst of deletions is equivalent to correcting a single burst of insertions.  Codes correcting a burst of exactly $b$ consecutive deletions, or a burst of up to $b$ consecutive deletions, were presented in \cite{SchWacGabYaa2017,LenPol2020}, with the redundancy being of optimal asymptotic order. The $b$-burst-correcting codes pertaining to deletions were treated in \cite{BitHanPolVor2021}, called codes correcting localized deletions therein, and a class of such codes of  asymptotically optimal redundancy was proposed. Similarly,  permutation codes correcting a single burst of  $b$ consecutive deletions were studied in \cite{CheLinNguVuWeiZha2020}.

This paper focuses on the model  of  \emph{limited-magnitude errors}, which  could  be found in several  applications, including high-density magnetic recording channels \cite{KuzVin93,LevVin93}, flash memories \cite{CasSchBohBru10}, and some DNA-based  storage systems \cite{JaiFarSchBru20,WeiWanSch21}. In all of these applications, information  is encoded as a  vector of integers $\vx \in \Z^n$. A $(\kp,\km)$-limited-magnitude error affects a position by increasing it  by as much as $\kp$ or decreasing it by as much as $\km$. The design of codes combating random limited-magnitude errors  has been extensively researched, see  e.g., \cite{Ste84,HamSte84,HicSte86,Ste90,SteSza94,KloLuoNayYar11,YarKloBos13,Sch14,ZhaGe16,ZhaZhaGe17,ZhaGe18,YeZhaZhaGe20,BuzEtz12,WeiWanSch21}. However, the applications which exhibit limited-magnitude errors are prone to errors occurring in a burst. The coding schemes for magnetic recording channels~\cite{KuzVin93,LevVin93}, and the DNA-based storage system of~\cite{JaiFarSchBru20}, all employ a constrained code as part of the system. Decoders for constrained codes are usually finite state machines, and an error in their decoding process causes a burst of errors in their output (e.g., see~\cite[Section~5.5]{LinMar85}). Similarly, flash memories suffer from inter-cell interference~\cite{EitRoy99}, leading again to bursts of errors. To the extent of our knowledge,  there is no research in the literature on codes correcting a single burst of limited-magnitude errors. We therefore focus in this paper on such codes, and in particular, perfect codes.

Following the research on bursts of substitutions, e.g., \cite{AbdMcEOdlTil86,Abd88,Etz01b},
we distinguish between \emph{cyclic bursts} and \emph{non-cyclic bursts}, of limited-magnitude errors. In the examples mentioned here, \cite{AbdMcEOdlTil86,Abd88} study cyclic bursts, whereas~\cite{Etz01b} studies non-cyclic bursts. We follow suit, and study both types of bursts. If a word $\vx\in \Z^n$ suffers a cyclic burst of length $b$, then we can write the corrupted vector as  $\vx+\ve$ for some $\ve$ in the error ball
\begin{equation}\label{eq:defcyclicburst}
\begin{split}
\cE^{\circ}(n,b,\kp,\km)\eqdef \{(e_0,e_1,\ldots,e_{n-1}) \in [-\km,\kp]^n ~|~ & \textup{ there is an  $i\in \Z_n$ such that } e_\ell= 0  \\
& \textup{ for all }  \ell \in \Z_n \setminus \set{i,i+1,\ldots,i+b-1} \}.
\end{split}
\end{equation}
If $\vx$ suffers a non-cyclic burst of length $b$, then
the corrupted vector is $\vx+\ve$ for some $\ve$ in the error ball
\begin{equation}\label{eq:defburst}
\begin{split}
\cE(n,b,\kp,\km)\eqdef \{\ve=(e_1,e_2,\ldots,e_{n}) \in [-\km,\kp]^n ~|~ &\textup{ there is an  $i\in [1,n]$ such that }  e_\ell= 0  \\
& \textup{ for all }  \ell \in [1,n] \setminus [i,\min\set{n,i+b-1}] \}.
\end{split}
\end{equation}
Note that in the cyclic case we use $\Z_n$ to label the coordinates and the addition is done in $\Z_n$ (i.e., modulo $n$), while in the non-cyclic case we use the set $[1,n]$ to label the coordinates  and the addition is operated in $\Z$.

 The subject of interest for this paper is perfect  codes correcting a single burst of limited-magnitude errors. Our main contributions are:
\begin{enumerate}
\item
For each $n\geq 2$, we construct a perfect  code of length $n$ which can correct a non-cyclic $2$-burst of $(1,0)$-limited-magnitude errors.
\item
For each $n\equiv 1,4\pmod{6}$, we construct a perfect  code of length $n$ which can correct a cyclic $2$-burst of $(1,0)$-limited-magnitude errors.
\item
We present a generic construction based on finite fields for cyclic codes correcting a cyclic $b$-burst of $(\kp,\km)$-limited-magnitude errors. This construction requires a primitive element satisfying some conditions as the input. Combining this construction and the  approach in \cite{AbdMcEOdlTil86}, we show the existence of a  class of perfect cyclic $b$-burst-correcting codes for each  $(b,\kp,\km)\in\set{(2,1,0),(2,1,1),(3,1,0),(3,1,1)}$.
\end{enumerate}
The parameters of the code constructions are summarized in Table~\ref{tab:summary}. We note that all the codes presented in this paper are lattice codes.

The paper is organized as follows. We begin, in Section~\ref{sec:prelim}, by providing notation and basic known results used throughout the paper. Section~\ref{sec:explcon} is devoted to the constructions of perfect codes correcting a $2$-burst of $(1,0)$-limited-magnitude errors. Both non-cyclic bursts and cyclic bursts are considered. Section~\ref{sec:confield} presents the generic construction for  codes correcting a single cyclic $b$-burst, and uses it to treat the cases of $(b,\kp,\km)\in\set{(2,1,0),(2,1,1),(3,1,0),(3,1,1)}$. In
Section~\ref{sec:comments} we summarize the results, and comment on extensions and open questions.

\section{Preliminaries}\label{sec:prelim}

For integers $a\leq b$ we define $[a,b]\eqdef\set*{a,a+1,\dots,b}$. For a sequence $\vs$, we use $\vs[i,j]$ to denote the subsequence of $\vs$ which starts at the position $i$ and ends at the position $j$.  We use $\Z_m$ to denote the
cyclic group of integers with addition modulo $m$, and $\F_q$ to
denote the finite field of size $q$.

We say $\cB\subseteq\Z^n$ \emph{packs} $\Z^n$ by $T\subseteq\Z^n$, if
the translates of $\cB$ by elements from $T$ do not intersect,
namely, for all $\vv,\vv'\in T$, $\vv\neq\vv'$,
\[ (\vv+\cB)\cap(\vv'+\cB)=\varnothing.\]
We say $\cB$ \emph{covers} $\Z^n$ by $T$ if
\[ \bigcup_{\vv\in T} (\vv+\cB) = \Z^n.\]
If $\cB$ both packs and covers $\Z^n$ by $T$, then we say that $\cB$
\emph{tiles} $\Z^n$ by $T$.
It now follows that a perfect code capable of correcting a cyclic burst in our setting is  equivalent to a tiling of $\Z^n$ by $\cE^\circ(n,b,\kp,\km)$ defined in \eqref{eq:defcyclicburst}, and a perfect code capable of correcting a non-cyclic burst in our setting is equivalent to a tiling of $\Z^n$ by $\cE(n,b,\kp,\km)$ defined in \eqref{eq:defburst}.

A code $\Lambda\subseteq\Z^n$ is called a \emph{lattice   code} if it  is an additive subgroup of
$\Z^n$. Similarly, we have the notion of lattice tilings.
Throughout the paper, we shall only consider lattice codes, since these are easier to
analyze, construct, and encode, than non-lattice codes.

\subsection{Group Splitting}
Perfect lattice codes that correct a single $(\kp,\km)$-limited-magnitude error are equivalent to lattice tilings of $\Z^n$ with $\cE(n,1,\kp,\km)$. If we treat each   point of  $\Z^n$ as a unit cube centered at it, then the shape   $\cE(n,1,\kp,\km)$ is called a \emph{cross} when $\kp=\km$, a \emph{semi-cross} when $\km=0$, and a \emph{quasi-cross} when $\kp\geq\km\geq 0$. The study of lattice tilings with these shapes  can be traced back to 1960's (e.g., see
\cite{Ste67}), and is usually connected with group splitting  (e.g.,
\cite{Ste84,HamSte84,HicSte86,Sch12,Sch14}). For an excellent
treatment and history, the reader is referred to~\cite{SteSza94} and
the many references therein.
 More recent results may be found
in~\cite{YeZhaZhaGe20} and the references therein.

To construct  codes that correct multiple  errors, the  notion of group splitting was generalized  in \cite{BuzEtz12}. Lattice tilings of chairs, or equivalently perfect lattice
codes that correct $n-1$ random $(\kp,0)$-limited-magnitude errors, were constructed there. Additionally, several non-existence results for perfect codes that correct multiple random errors can be found in \cite{BuzEtz12,WeiSchwartz}.
In this paper, we shall study lattice codes that correct a single burst of limited-magnitude errors by using the concept of (generalized) group splitting.

Let $G$ be a finite Abelian group, where $+$ denotes the group
  operation. For $m\in\Z$ and $g\in G$, let $mg$ denote $g+g+\dots+g$
  (with $m$ copies of $g$) when $m>0$, which is extended in the
  natural way to $m\leq 0$. For  a sequence $\vm=(m_1,m_2,\ldots,m_n)\in \Z^n$ and a  sequence $\vs=(s_1,s_2,\ldots,s_n)\in G^n$, we denote
  \[\vm \cdot \vs \eqdef \sum_{i=1}^m m_i s_i . \]

\begin{definition}
  \label{def:generalsplit}
  A set $\cA \subset \Z^n$ \emph{splits} an Abelian group $G$ with a \emph{splitting sequence} $\vs=(s_1,s_2,\ldots,s_n) \in G^n$ if the set $\set{\va\cdot \vs; \va\in \cA}$ contains $\abs{\cA}$ distinct  elements of $G$. This operation is called  a \emph{(generalized) splitting}.
\end{definition}

In our context of $b$-burst-correcting codes with respect to $(\kp,\km)$-limited-magnitude errors,  we need to take $\cA=\BURB$ or $\cA=\CBURB$. The following  theorems show the equivalence of lattice tiling of  $\Z^n$ and splitting.

\begin{theorem}[Lemma 4 and Corollary 1 in \cite{BuzEtz12}]
  \label{th:lattotile}
  Let $\cS\subset\Z^n$ be a finite subset, and  $G$  be an Abelian group of order $\abs{\cS}$. Assume that $\cS$ splits $G$ with a splitting sequence $\vs$. Define $\phi:\Z^n\to G$ as
  $\phi(\vx)\eqdef\vx\cdot(s_1,\dots,s_n)$ and let
  $\Lambda\eqdef\ker\phi$. Then $\Lambda$ is a lattice tiling of
  $\Z^n$ with  $\cS$.
\end{theorem}

\begin{theorem}[Lemma 3 and Corollary 1 in \cite{BuzEtz12}]
  \label{th:tiletolat}
  Let $\Lambda\subseteq\Z^n$ be a lattice tiling of $\Z^n$ with  $\cS\subset\Z^n$, and assume both $\cS$ and $G$ are finite. Define $G\eqdef \Z^n / \Lambda$. Let $\phi:\Z^n\to G$ be the natural homomorphism, namely
  the one that maps any $\vx\in\Z^n$ to the coset of $\Lambda$ in
  which it resides. Set  $\vs\eqdef (\phi(\ve_1), \phi(\ve_2), \ldots, \phi(\ve_n)  )$,  where $\ve_i$ is
 the $i$-th unit vector in $\Z^n$. Then $\cS$ splits $G$ with the splitting sequence $\vs$.
\end{theorem}

Splittings with $\cE(n,b,\kp,\km)$ can also be used to characterize codes that correct a single burst of substitutions. Let $p$ be a prime and let $\kp$ and $\km$ be non-negative integers such that $\kp+\km+1=p$. Let  $\cC$ be an $[n,n-r]_p$-linear code with parity-check matrix
$H$. We treat the columns of $H$ as elements of $\F_p^r$ and denote them as $h_1,h_2,\ldots,h_n$. Then $\cC$ is a perfect $b$-burst-correcting code with respect to  substitutions if and only if $p^r=\abs{\cE(n,b,\kp,\km)}$ and the additive group $\F_p^r$ can be split by $\cE(n,b,\kp,\km)$ with the sequence $\vh=(h_1,h_2,\ldots,h_n)$.
Binary perfect $2$-burst-correcting codes pertaining to substitutions were studied in \cite{Etz01b} and  a  construction for  their parity-check matrices was presented.  The existence result of such codes could be stated as follows in the language of splittings.

\begin{theorem}[\cite{Etz01b}]
For each $r\geq 5$, there exists a splitting of $\F_2^r$ by $\cE(2^{r-1},2,1,0)$.
\end{theorem}

In the following two sections we are going to present some other constructions of splittings by  $\BURB$ or $\CBURB$. These tilings are equivalent to perfect $b$-burst-correcting codes with respect to limited-magnitude errors, by taking the kernel of the map $\phi(x)$ defined in Theorem~\ref{th:lattotile}.

\section{Perfect $2$-Burst-Correcting Codes for $(1,0)$-Limited-Magnitude Errors}
%\section{Lattice Tilings of $\cE(n,2,1,0)$ and $\cE^{\circ}(n,2,1,0)$}
\label{sec:explcon}

In this section, we present a class of constructions for $2$-burst-correcting codes with $(1,0)$-limited-magnitude errors, both for cyclic bursts as well as for non-cyclic bursts. Our constructions are based on splitting the cyclic group $\Z_g$. Using these constructions, together with Theorem~\ref{th:lattotile}, we show that $\Z^n$ can be lattice tiled by $\cE(n,2,1,0)$ for all $n \geq 2$, and that $\Z^n$ can be lattice tiled by $\cE^\circ(n,t,1,0)$ for all $n \equiv 1,4 \pmod{6}$.

The basic idea behind these constructions comes from design theory: we start with a short sequence $(a_1,a_2,\ldots,a_s)$ that satisfies a certain property, and develop it by adding a series of numbers $(0,b,2b,\ldots,tb)$ to each element $a_i$. In this way, we obtain a long sequence
\[(a_1,a_2,\ldots,a_s, a_1+b,a_2+b, \ldots,a_s+b, \ldots,a_1+tb,a_2+tb,\ldots, a_{i_0}+tb)\]
for some $1\leq i_0\leq s$, which is usually the desired splitting sequence. We note that   $\set{0,b,2b,\ldots,tb}$ need not form a subgroup of $\Z_g$.

Since the operation described above repeats throughout our construction, we introduce the following succinct notation. Let $\va=(a_1,a_2,\dots,a_n)\in\Z_g^n$ and $\vb=(b_1,b_2,\dots,b_m)\in\Z_g^m$ be two vectors, not necessarily of the same length. We define
\begin{align*}
\va\boxplus \vb & \eqdef \One_m\otimes \va + \vb\otimes\One_n \\
& = (a_1+b_1,a_2+b_1,\dots,a_n+b_1,a_2+b_1,a_2+b_2,\dots,a_n+b_2, \dots, a_1+b_m, a_2+b_m,\dots, a_n+b_m),
\end{align*}
where $\otimes$ denotes the Kronecker product, and $\One_\ell$ denotes a row vector of all ones with length $\ell$. If we wish to keep only the first $\ell$ entries of $\va\boxplus\vb$ we shall use the notation we have already defined, $(\va\boxplus\vb)[1,\ell]$.

We first give our constructions in the case of non-cyclic bursts. In this case, we have $\abs{\cE(n,2,1,0)}=2n$, and we are going to the split the  group $\Z_{2n}$ by $\cE(n,2,1,0)$.

\begin{theorem}
\label{thm:b=2kp=1km=0-nc}
Let $n\geq 2$. Then $\Z^n$ can be lattice tiled by $\cE(n,2,1,0)$. Namely, there exists a perfect lattice code in $\Z^n$ which can correct a single non-cyclic $2$-burst of  $(1,0)$-limited-magnitude errors.
\end{theorem}

\begin{IEEEproof}
The proof proceeds by considering four cases, depending on the residue of $n$ modulo $4$.

\textbf{Case 1:} Assume $n=2m+1$ where $m\geq 1$ is even. Working in the group $G=\Z_{4m+2}$, let us define
\begin{align*}
\vs&\eqdef \parenv*{(m+1,3m+3)\boxplus (0,2,4,\ldots,2m)}[1,n] \\
&=(m+1,3m+3, m+3, 3m+5, \ldots, \\
&\quad\ \  m+1+2(m-1)=3m-1, 3m+3+2(m-1)=m-1, m+1+2m=3m+1 ).
\end{align*}
Note that
 \[\set*{\vs[i]; 1\leq i\leq n}=\set*{m+1,m+3,\ldots,3m-1, 3m+1,3m+3,3m+5,\ldots,m-1}=\set*{1,3,5,\ldots,4m+1}\]
 and
 \[ \set*{\vs[i]+\vs[i+1]; 1\leq i \leq n-1  } = \set*{2,4,6,\ldots,4m}.\]
Thus, $G$ is split by $\cE(n,2,1,0)$ with $\vs$.

\textbf{Case 2:} Assume $n=2m+1$ where $m\geq 1$ is odd. Once again we work in $G=\Z_{4m+2}$, but now we define
\begin{align*}
\vs&\eqdef\parenv*{(3m+2,m+2)\boxplus (0,2,4,\ldots,2m)}[1,n]\\
&=(3m+2 ,m+2, 3m+4, m+4, \ldots, \\
&\quad\ \ 3m+2+2(m-1)=m-2, m+2+2(m-1)=3m, 3m+2+2m=m ).
\end{align*}
Then
\[\set*{\vs[i]; 1\leq i\leq n}=\set*{3m+2,3m+4,\ldots,m-2,m, m+2,m+4,\ldots, 3m}=\set*{1,3,5,\ldots,4m+1}\]
and
\[ \set*{\vs[i]+\vs[i+1]; 1\leq i \leq n-1  } = \set*{2,4,6,\ldots,4m}.\]
Hence, $G$ is split by   $\cE(n,2,1,0)$ with $\vs$.

\textbf{Case 3:} Assume $n=2m$ where $m\geq 1$ is even. This time we work in $G=\Z_{4m}$, and we define
\begin{align*}
\vs &\eqdef (m+1,3m+1) \boxplus (0,2,4,\ldots,2(m-1)) \\
&=(m+1,3m+1, m+3, 3m+3, \ldots, m+1+2(m-1)=3m-1, 3m+1+2(m-1)=m-1).
\end{align*}
Then
\[\set*{\vs[i]; 1\leq i\leq n}=\set*{m+1,m+3,\ldots,3m-1,3m+1,3m+3,\ldots,m-1}=\set*{1,3,5,\ldots,4m-1}\]
and
\[ \set*{\vs[i]+\vs[i+1]; 1\leq i \leq n-1  } = \set*{2,4,6,\ldots,4m-2}.\]
It follows that $G$ is split by  $\cE(n,2,1,0)$ with $\vs$.

\textbf{Case 4:} Assume $n=2m$ where $m\geq 1$ is odd. We again work in $G=\Z_{4m}$, but this time the splitting is more involved. For $m=1$, it is easily seen that $\Z_{4}$ is split by  $\cE(2,2,1,0)$ with the splitting sequence $(1,2)$.
For $m\geq 3$, consider the following sequences
\begin{align*}
\vs_1 & \eqdef (1,3,5,\ldots,2m-3),\\
\vs_2 & \eqdef (2m+1, 2m+5,2m+9,\ldots,4m-1),\\
\vs_3 & \eqdef (4m-3, 4m-7,4m-11, \ldots, 2m-1).
\end{align*}
Denote
\[\vs\eqdef \vs_1\vs_2\vs_3.\]
Then $\vs$ has length $n=2m$ and
\[\set*{\vs[i]; 1\leq i \leq 2m}= \set*{1,3,5,\ldots,4m-1}.\]
We have that
\begin{align*}
\set*{\vs_1[i]+\vs_1[i+1]; 1\leq i \leq m-2} &= \set{4,8,12,\ldots,4m-8}, \\
\set*{\vs_2[i]+\vs_2[i+1]; 1\leq i \leq \frac{m-1}{2}} &= \set{6,14,22,\ldots,4m-6}, \\
\set*{\vs_3[i]+\vs_3[i+1]; 1\leq i \leq \frac{m-1}{2}} &= \set{2,10,18,\ldots,4m-10}.
\end{align*}
Additionally,
\begin{align*}
\vs_1[m-1]+\vs_2[1]&=2m-3+2m+1=4m-2 \\
\vs_2\sparenv*{\frac{m+1}{2}}+\vs_3[1]&=4m-1+4m-3=4m-4.
\end{align*}
It follows that
\[\set*{\vs[i]+\vs[i+1]; 1\leq i \leq 2m-1}=\set{2,4,\ldots,4m-2}.\]
Thus, $G$ is split by  $\cE(n,2,1,0)$ with $\vs$.
\end{IEEEproof}

We now move to the case of cyclic bursts. In this case, we have $\abs{\cE^{\circ}(n,2,1,0)}=2n+1$, and so we consider the group $\Z_{2n+1}$.

\begin{theorem}
\label{thm:b=2kp=1km=0-c2}
Let $n\geq 4$ be a positive integer such that $n\equiv 1,4\pmod{6}$. Then $\Z^n$ can be lattice tiled by $\cE^{\circ}(n,2,1,0)$. Namely, there exists a perfect lattice code in $\Z^n$ which can correct a single cyclic $2$-burst of $(1,0)$-limited-magnitude errors.
\end{theorem}

\begin{IEEEproof}
We divide our proof depending on the residue $n$ leaves modulo $6$.

\textbf{Case 1:} Assume that $n=6m+1$, $m\geq 1$. We work in the group $G=\Z_{12m+3}$ and show that it can be split by $\cE^{\circ}(n,2,1,0)$. Let
\[a=3m+1,b=3m+2,c=6m+2,d=6m+4,e=2,f=9m+5,\]
and define
\begin{align*}
\vs&\eqdef ((a,b,c,d,e,f)\boxplus(0,3,6,\ldots,3m))[1,n] \\
&=(a,b,\ldots ,f,a+3,b+3,\ldots,f+3, \ldots,a+3(m-1),b+3(m-1),\ldots,f+3(m-1), a+3m).
\end{align*}
We now observe that
\begin{align*}
\set*{a+3i ; 0\leq i \leq m } \cup \parenv*{\bigcup_{i=0}^{m-1} \set*{d+3i,b+c+6i,e+f+6i}}&=\set*{1,4,7,\ldots,12m+1}, \\
\parenv*{\bigcup_{i=0}^{m-1} \set*{b+3i,c+3i,e+3i,f+3i}}\cup\set*{2a+3m}&=\set{2,5,8,\ldots,12m+2},\\
\bigcup_{i=0}^{m-1} \set*{a+b+6i,c+d+6i,d+e+6i,f+a+3+6i}&=\set*{3,6,9,\ldots,12m}. \end{align*}
Hence, $G$ is split by  $\cE^\circ(n,2,1,0)$ with $\vs$.

\textbf{Case 2:} Assume that $n=6m+4$. We now work in the group $G=\Z_{12m+9}$ and show that it can be split by $\cE^{\circ}(n,2,1,0)$. For $m=0$, it is easily seen that $\Z_{9}$ is split by  $\cE^{\circ}(4,2,1,0)$ with the splitting sequence $(1,3,2,6)$. For $m\geq 1$,  let \[a=1,b=9m+10,c=3m+2,d=3m+7,e=6m+7,f=6m+8,\]
and define
\begin{align*}
\vs_1&\eqdef (a,b,c,d,e,f)\boxplus (0,3,6,\ldots,3(m-1))\\
&=(a,b,\ldots ,f,a+3,b+3,\ldots,f+3, \ldots,a+3(m-1),b+3(m-1),\ldots,f+3(m-1)), \\
\vs_2&\eqdef (6m+5,12m+6,6m+6,9m+7).
\end{align*}
Define $\vs$ to be the concatenation of $\vs_1$ and $\vs_2$, i.e.,
\[\vs\eqdef\vs_1\vs_2.\]
Note that
\begin{align*}
\bigcup_{i=0}^{m-1} \set*{a+3i,b+3i,d+3i,e+3i}&=\set*{1,4,7,\ldots,12m+7} \setminus \set*{3m+1,3m+4,9m+7}, \\
\bigcup_{i=0}^{m-1} \set*{c+3i,f+3i,a+b+6i,d+e+6i}&=\set{2,5,8,\ldots,12m+8} \setminus \set*{6m+2,6m+5,9m+8},
\end{align*}
and
\begin{align*}
& \ \parenv*{\bigcup_{i=0}^{m-1} \set*{b+c+6i,c+d+6i,e+f+6i}} \cup  \set*{f+a+3+6i; 0\leq i\leq m-2} \\  = & \ \set{3,6,9,\ldots,12m+3} \setminus\set*{6m+3,6m+6}.
\end{align*}
Furthermore, we have
\[\set*{\vs_2[i];1\leq i\leq 4}=\set*{6m+5,12m+6,6m+6,9m+7}\]
and
\begin{align*}
    & \ \set*{f+3(m-1)+\vs_2[1],\vs_2[1]+\vs_2[2],\vs_2[2]+\vs_2[3], \vs_2[3]+\vs_2[4],\vs_2[4]+a}\\
    = & \ \set*{3m+1,6m+2,6m+3,3m+4,9m+8}.
\end{align*}
Hence, $G$ is split by  $\cE^\circ(n,2,1,0)$ with $\vs$.
\end{IEEEproof}

\section{Perfect $\leq 3$-Cyclic-Burst-Correcting Codes for $(1,1)$ and $(1,0)$-Limited-Magnitude Errors}
\label{sec:confield}

In this section, we present a construction  for  the splitting of the additive group of $\F_q$ by  $\cE^{\circ}(n,t,\kp,\km)$. Thus, throughout this section, we let $G$ be the additive group of $\F_q$. This is in contrast with the previous section, where we split only cyclic groups. Denote
\begin{equation}
    \label{eq:defe}
e\eqdef(\kp+\km)(\kp+\km+1)^{b-1}.
\end{equation}
Let $q$ be a prime power such that $e|q-1$, and denote
\begin{equation}
    \label{eq:defn}
n\eqdef (q-1)/e.
\end{equation}
Then
\begin{equation}
    \label{eq:cycball}
\abs{\CBURB}=en+1=q.
\end{equation}

Let $\alpha \in \F_q^*$ be a primitive element. For any $z\in\F^*_q$, we use $\ord_\alpha(z)$ to denote the  unique integer $a\in [0,q-2]$ such that $z=\alpha^{a}$.

The splitting sequence we shall use most of this section is
defined as
\[\vs_\alpha\eqdef (\alpha^{0}, \alpha^{e}, \alpha^{2e}, \ldots, \alpha^{(n-1)e}).\]
We also define
\begin{equation}
    \label{eq:deff}
\cF_b^{\kp,\km} \eqdef \set*{ (1,x^{e},x^{2e}, \ldots,x^{(b-1)e}) \cdot \vc ; \vc = (c_0,c_1,\ldots,c_{b-1}) \in [-\km,\kp]^b \textup{ and } c_0\not=0 }.
\end{equation}
Hence, $\cF_b^{\kp,\km}$ is a set of $e$ polynomials. The following result shows that by carefully choosing  $\alpha$, the group $G$ can be split  by $\CBURB$   with  $\vs_\alpha$.

\begin{proposition}\label{prop:fieldcon}
Assume the setting above, and $n\geq 2b-1$. Let $\alpha$ be a primitive element of $\F_q^*$, and assume $f(\alpha)\neq 0$ for all $f(x)\in\cF_b^{\kp,\km}$. If
\begin{equation}\label{eq:condition1}
\set*{\ord_\alpha(f(\alpha)) \ppmod{e} ; f(x) \in \cF_b^{\kp,\km} } = \set*{0,1,2,\ldots,e-1},
\end{equation}
then $\CBURB$ splits $G$ (the additive group of $\F_q$) with the splitting sequence $\vs_\alpha$.
\end{proposition}

\begin{IEEEproof}
For each  vector $\vc=(c_0,c_1,\ldots,c_{b-1})\in [-\km,\kp]^b$, let
\[\cE_\vc \eqdef \set*{\ve=(e_0,e_1,\ldots,e_{n-1}) \in \CBURB; \textup{ there is an integer $i$ such that }   \ve[i,i+b-1]=\vc },\]
where the indices of $\ve$ are taken cyclically, i.e., modulo $n$. Since $n\geq 2b-1$, it follows that $\CBURB\setminus \set{{\bf 0}}$ can be partitioned into $\cE_{\vc}$'s, where $\vc=(c_0,c_1,\ldots,c_{b-1}) \in [-\km,\kp]^{b}$ and $c_0\neq0$.

Note that
\begin{align*}
\set*{\ve \cdot \vs_\alpha ; \ve \in \cE_{\vc} } & = \set*{ \alpha^{e\ell}(c_0+c_1\alpha^{e}+c_2\alpha^{2e}+\cdots+c_{b-1}\alpha^{(b-1)e}) ; \ell \in [0,n-1]}\\
& =\set*{ \alpha^{e\ell+a}; \ell \in [0,n-1]  },
\end{align*}
where $a = \ord_\alpha(c_0+c_1\alpha^{e}+c_2\alpha^{2e}+\cdots+c_{b-1}\alpha^{(b-1)e})$.

Since \eqref{eq:condition1} holds, the collection of sets of the form $\set*{\ve \cdot \vs_\alpha ; \ve \in \cE_{\vc} }$, where $\vc \in [-\km,\kp]^{b}$ with $c_0\neq0$, are exactly the $e$ cosets of the multiplicative subgroup $\langle \alpha^e \rangle$ in $\F_q^*$. It then follows that
\[\set*{\ve \cdot \vs_\alpha ; \ve \in \CBURB } = \set{0} \cup \parenv*{ \bigcup_{\substack{\vc \in [-\km,\kp]^{b}\\ c_0\neq 0}} \set*{\ve \cdot \vs_\alpha ; \ve \in \cE_{\vc} } }=\F_q.\]
Hence, in conjunction with~\eqref{eq:cycball}, $\CBURB$ splits $G$ with $\vs_\alpha$.
\end{IEEEproof}

According to  Theorem~\ref{th:lattotile}, the splitting in Proposition~\ref{prop:fieldcon} yields  a lattice tiling of $\Z^n$ by $\CBURB$, or equivalently, a perfect lattice code which can correct a cyclic $b$-burst of $(\kp,\km)$-limited-magnitude errors. Furthermore, noting that $(x_0,x_1,\ldots,x_{n-1})\cdot \vs_\alpha=0$ implies that $(x_{n-1},x_0,\ldots,x_{n-2})\cdot \vs_\alpha= \alpha^e \cdot((x_0,x_1,\ldots,x_{n-1})\cdot \vs_\alpha) =0$, the code itself is cyclic.

Let us start examining specific values of the code parameters. When $b=2$ and $(\kp,\km)=(1,0)$, we have $e=2$ and $q$ is odd. As we shall soon observe and use, the sufficient condition~\eqref{eq:condition1} is reduced to that of $1+\alpha^2$ being a quadratic non-residue. Since any primitive element of $\F_q$, $q\geq 3$, is always a quadratic non-residue, the following result can be used for our construction.

\begin{lemma}[{{\cite[Theorem~1]{Bookeretal2019}}}]
\label{lm:1nasquare}
Let $q$ be an odd prime power which does not belong to the following set:
\begin{equation}
\label{eq:defE}
E\eqdef \set{3, 5, 9, 7, 11, 13, 19, 23, 25, 29, 31, 37, 41, 43, 49, 61, 67, 71, 73, 79, 121, 127, 151, 211}.
\end{equation}
Then there is a primitive element $\alpha \in \F_q$ such that $1+\alpha^2$ is also a primitive element of $\F_q$.
\end{lemma}

\begin{theorem}\label{thm:b=2kp=1km=0-c1}
Let $q\geq 7$ be an odd prime power, and let $n=(q-1)/2$. Then there is a perfect lattice code of $\Z^n$ which can correct a single cyclic $2$-burst of $(1,0)$-limited-magnitude errors.
\end{theorem}

\begin{IEEEproof}
For $q=7$, let $G=\Z_7$ and  $\vs=(1,2,4)$. Then $\abs{G}=\abs{\cE^\circ(3,2,1,0)}$ and $G$ is split by $\cE^\circ(3,2,1,0)$ with $\vs$. According to Theorem~\ref{th:lattotile}, there is a lattice tiling of $\Z^3$ by $\cE^\circ(3,2,1,0)$. This specific case is in fact a standard $2$-error-correcting code, and since $n=3$, it is a perfect tiling with a chair~\cite{BuzEtz12}.

For $q \geq 9$, let $G$ be the additive group of $\F_q$. With the parameters of this theorem, we have
\[ \cF_{2}^{1,0}=\set*{1,1+x^2}.\]
We would like to use Proposition~\ref{prop:fieldcon} to construct the splitting. Since $\ord_{\alpha}(1)=0$, we need $\ord(1+\alpha^2)\equiv 1 \pmod{2}$, namely, that $1+\alpha^2$ is a quadratic non-residue. If $q \not \in E$ of~\eqref{eq:defE}, then Lemma~\ref{lm:1nasquare}
shows that there is a primitive $\alpha$ such that $1+\alpha^2$ is also primitive, and hence, $1+\alpha^2$ is a quadratic non-residue. If  $q \in E$ and $q\geq 9$, a computer search shows that there is a primitive element $\alpha\in \F_q$ with $1+\alpha^2$ being a quadratic non-residue. According to   Proposition~\ref{prop:fieldcon}, $\cE^\circ(n,2,1,0)$ splits   $G$ with $\vs_\alpha$.   The conclusion then follows from  Theorem~\ref{th:lattotile} and the fact that $\abs{G}=\abs{\cE^\circ(n,2,1,0)}$.
\end{IEEEproof}

We note that both   Theorem~\ref{thm:b=2kp=1km=0-c2} and Theorem~\ref{thm:b=2kp=1km=0-c1} concern the tiling of the ball $\cE^{\circ}(n,2,1,0)$, but in different regimes. In Theorem~\ref{thm:b=2kp=1km=0-c1} the size $\abs{\cE^{\circ}(n,2,1,0)}$ is $q$, a prime power, while  in Theorem~\ref{thm:b=2kp=1km=0-c2} the size $\abs{\cE^{\circ}(n,2,1,0)}$ is divisible by $3$.

For the other cases, we adapt the approach in  \cite{AbdMcEOdlTil86} to show  the existence of  $\alpha$ which satisfies condition \eqref{eq:condition1}. Recall that a \emph{multiplicative character} of  $\F^*_q$ is a group homomorphism $\chi:\F_q^*\to\C$, such that for all $\beta,\gamma\in \F_q^*$ we have $\chi(\beta\gamma)=\chi(\beta)\chi(\gamma)$. We use $\chi^i(\beta)=(\chi(\beta))^i$ to avoid awkward parentheses, hence the superscript $i$ denotes taking the $i$th power of $\chi(\beta)$ and not function composition. We say that $\chi$ has order $i$ if $i$ is the minimal positive integer such that $\chi^i(\beta)=1$ for all $\beta\in \F_q^*$. Thus, the order of $\chi$ divides $q-1$. Let $\chi_i$  denote an arbitrary multiplicative character  of order $i$. In particular, $\chi_1$ is the function sending all the elements of $\F_q^*$ to $1$. It is convenient to extend the definition by letting $\chi(0)=0$ for all characters. We also recall the definition of the M\"obius function, $\mu:\N\to\set{-1,0,1}$. If $n\in\N$ is a natural number, $n=\prod_{i=1}^{s}p_i^{m_i}$, where $m_i\in\N$ and the $p_i$ are distinct primes, then
\[ \mu(n) = \begin{cases}
0 & \text{$m_i\geq 2$ for some $i$,}\\
(-1)^s & \text{otherwise.}
\end{cases}
\]
The following sequence of lemmas will help us establish the existence of perfect codes.

\begin{lemma}[{{\cite[Lemma~4]{Carlitz1952}}}]
\label{lm:primitive}
For all $\alpha\in\F^*_q$ we define
\begin{equation}\label{eq:defpsi}
\psi(\alpha)\eqdef \sum_{k \mid q-1} \frac{\mu(k)}{k} \sum_{\chi^k=\chi_1 }\chi(\alpha),
\end{equation}
where $\mu$ is the M\"obius function. Then
\begin{equation*}
\begin{split}
  \psi(\alpha) = & \begin{cases}
    1 & \text{if $\alpha$ is primitive,} \\
    0 & \text{otherwise.}
\end{cases}
\end{split}
\end{equation*}
\end{lemma}

In the following we also use the convention that $0^0=0$ to simplify derivations.

\begin{lemma}
\label{lem:theta}
Assume the setting above. Furthermore, let $\cF=\set{f_1,f_2,\ldots, f_{M}}\subseteq \F_q[x]$ be a collection of polynomials over $\F_q$, and let $h$ be an integer such that $h|q-1$. For any $\alpha\in\F^*_q$ we define
\begin{equation}\label{eq:deftheta}
\Theta(\alpha)\eqdef \psi(\alpha) \prod_{i=1}^{M} \sum_{j=0}^{h-1} \chi_h^j (\alpha^{-\ell_i} f_i(\alpha)),
\end{equation}
where $\ell_i\in\Z$ for all $i$. Then
\[
\Theta(\alpha) =  \begin{cases}
    h^{M} & \text{if $\alpha$ is primitive and for all $1\leq i \leq M$, $f_i(\alpha)\neq 0$, $\ord_\alpha(f_i(\alpha))\equiv \ell_i \ppmod{h}$} \\
    0 & \text{ otherwise.}
\end{cases}
\]
\end{lemma}

\begin{IEEEproof}
If $\alpha$ is not primitive, then by Lemma~\ref{lm:primitive}, $\psi(\alpha)=0$, and therefore also $\Theta(\alpha)=0$. If $f_i(\alpha)=0$ for some $i$, then again, $\Theta(\alpha)=0$. We are therefore left with the case that $\alpha$ is primitive, and $f_i(\alpha)\neq 0$ for all $i$. Let $\gamma\in\F^*_q$, and assume $\ord_\alpha(\gamma)=m$. Then
\[ \chi_h(\gamma)=\chi_h(\alpha^m)=\chi_h^m(\alpha).\]
Thus, $\chi_h(\gamma)=1$ if and only if $m=\ord_{\alpha}(\gamma)\equiv 0\pmod{h}$. If indeed $\chi_h(\gamma)=1$, then
\[ \sum_{j=0}^{h-1} \chi_h^j(\gamma) = \sum_{j=0}^{h-1} 1 = h.\]
Otherwise, $\chi_h(\gamma)\neq 1$ and we have
\[ \sum_{j=0}^{h-1} \chi_h^j(\gamma) = \frac{\chi_h^h(\gamma) -1}{\chi_h(\gamma)-1} = 0.\]
Using this observation we note that
\[
 \sum_{j=0}^{h-1} \chi_h^j (\alpha^{-\ell_i} f_i(\alpha)) =  \begin{cases}
 h & \text{if $\ord_\alpha(f_i(\alpha))\equiv \ell_i \ppmod{h}$,} \\
 0 & \text{otherwise.}
 \end{cases}
\]
The conclusion now easily follows.
\end{IEEEproof}

Since we are working with characters, we shall also need a bound on character sums over $\F_q^*$, which can be derived from the Weil bound (see \cite{AbdMcEOdlTil86}).

\begin{lemma}[\cite{AbdMcEOdlTil86}]
\label{lm:weilsum-m}
Let $\chi$ be a multiplicative character of order $m>1$, and let $f\in \F_q[x]$ be a  polynomial  that cannot be written as $c\cdot (h(x))^m$ for any  $c\in\F_q$ and $h(x) \in \F_q[x]$. Let $d$ be the number of distinct roots of $f$ in its splitting field. Then for every $a\in \F_q$ we have
\[ \abs*{ \sum_{x\in \F_q^*} \chi(af(x) )} \leq  d\sqrt{q}.\]
\end{lemma}

For the next lemma we recall the definitions of Euler's function $\phi(n)$ and the divisor function $d(n)$, for all $n\in\N$,
\begin{align*}
\phi(n) & \eqdef \abs*{\set*{1\leq i\leq n ; \gcd(i,n)=1}}, \\
d(n) &\eqdef \sum_{i|n} 1.
\end{align*}

\begin{lemma}
\label{lm:existence}
Consider the setting of Lemma~\ref{lem:theta}. Suppose that for any $(i_1,i_2,\ldots,i_{M}) \in [0,h-1]^{M} \setminus \set{(0,0,\ldots,0)}$, the polynomial  $\prod_{t=1}^{M} (f_t(x))^{(q-1)i_t/h}$ cannot be written in the form $c\cdot (h(x))^{q-1}$, where $c\in\F_q$ and $h(x) \in \F_q[x]$. Then
\[\abs*{ \sum_{\alpha \in \F_q^*} \Theta(\alpha) -\phi(q-1)  } \leq A\cdot d(q-1)\cdot \sqrt{q},\]
where $\phi$ is the Euler function, $d$ is the divisor function, and $A$ is independent of $q$.
\end{lemma}

%We need to require that $\prod_{t=1}^{e-1} (f_t(x))^{(q-1)i_t/e}$ is not a $(q-1)$st power of a polynomial for any $(i_1,i_2,\ldots,i_{e-1}) \in [0,e-1]^{e-1} \setminus \set{(0,0,\ldots,0)}$.

\begin{IEEEproof}
From \eqref{eq:deftheta}, if $f_t(\alpha)\neq 0$ for all $1\leq t\leq M$, then let us write $\Theta(\alpha)=\psi(\alpha)+R(\alpha)$, where
\begin{equation}\label{eq:defR}
R(\alpha) = \psi(\alpha) \sum_{\substack{(i_1,i_2,\ldots,i_{M}) \in [0,h-1]^{M} \\ (i_1,i_2,\dots,i_M)\neq (0,0,\ldots,0)}}  \chi_h^{i_1}(\alpha^{-\ell_1}f_1(\alpha)) \cdots \chi_h^{i_M}(\alpha^{-\ell_M}f_M(\alpha)).
\end{equation}
Otherwise, if $f_t(\alpha)=0$ for some $1\leq t\leq M$, then $\Theta(\alpha)=0=R(\alpha)$. By Lemma~\ref{lm:primitive},
\[ \sum_{\alpha\in\F^*_q}\psi(\alpha) = \phi(q-1).\]
Thus, summing over all $\alpha\in \F_q^*$, we get
\begin{equation}
\label{eq:thetabelowabove}
\phi(q-1)-\sum_{t=1}^{M}{\deg(f_t)}+\sum_{\alpha \in \F_q^*}R(\alpha) \leq   \sum_{\alpha \in \F_q^*} \Theta(\alpha) \leq \phi(q-1)+\sum_{\alpha \in \F_q^*}R(\alpha).
\end{equation}
Note that $\sum_{t=1}^{M}{\deg(f_t)}$ is independent of $q$.
In the following, we shall give an upper bound on $\abs*{ \sum_{\alpha \in \F_q^*}R(\alpha)}$.

Let us observe a typical term in the sum on the right-hand side of~\eqref{eq:defR}. From \eqref{eq:defpsi}, we have
\[\psi(\alpha)  \chi_h^{i_1}(\alpha^{-\ell_1}f_1(\alpha)) \cdots \chi_h^{i_M}(\alpha^{-\ell_M}f_M(\alpha)) =   \sum_{k \mid q-1} \frac{\mu(k)}{k} \sum_{\chi^k=\chi_1 }\chi(\alpha)\chi_h^{i_1}(\alpha^{-\ell_1}f_1(\alpha)) \cdots \chi_h^{i_M}(\alpha^{-\ell_M}f_M(\alpha)).\]
In the inner sum, $\chi=\chi_j$ for some $j|k$. Hence
\[\chi_j(\alpha)\chi_h^{i_1}(\alpha^{-\ell_1}f_1(\alpha)) \cdots \chi_h^{i_M}(\alpha^{-\ell_M}f_M(\alpha))=\chi_{q-1}(\alpha^L w(\alpha))\]
where
\[L=(q-1)\parenv*{\frac{1}{j} -\frac{1}{h}(i_1\ell_1+\cdots+i_M\ell_M)  }\]
and
\[w(x)=\prod_{t=1}^M (f_t(x))^{(q-1)i_t/h}.\]
We notice that $w(x)$ has at most $\sum_{t=1}^M \deg f_t$ distinct roots in its splitting field. Due to the assumption, we can apply  Lemma~\ref{lm:weilsum-m} to  get
\[\abs*{\sum_{\alpha\in \F_q^*} \chi_j(\alpha)\chi_h^{i_1}(\alpha^{-\ell_1}f_1(\alpha)) \cdots \chi_h^{i_M}(\alpha^{-\ell_M}f_M(\alpha))} \leq \sum_{t=1}^M (\deg f_t) \sqrt{q}.  \]
Next, we observe that there are exactly $k$ characters $\chi$ such that $\chi^k=\chi_1$. Hence,
\[\abs*{\sum_{\alpha\in \F_q^*} \psi(\alpha)\chi_h^{i_1}(\alpha^{-\ell_1}f_1(\alpha)) \cdots \chi_h^{i_M}(\alpha^{-\ell_M}f_M(\alpha))} \leq \sum_{t=1}^M (\deg f_t) d(q-1) \sqrt{q}.  \]
It follows that
\[ \abs*{\sum_{\alpha\in \F_q^*} R(\alpha)} \leq \sum_{t=1}^M (h^M-1) (\deg f_t) d(q-1) \sqrt{q}.  \]
Finally, using~\eqref{eq:thetabelowabove}, we have
\begin{align*}
\abs*{ \sum_{\alpha \in \F_q^*} \Theta(\alpha) -\phi(q-1)  } & \leq \abs*{\sum_{\alpha\in \F_q^*} R(\alpha)} + \sum_{t=1}^M \deg f_t \leq \sum_{t=1}^M h^M (\deg f_t) d(q-1) \sqrt{q}.
\end{align*}
\end{IEEEproof}

Now, we study the cases   $(b,\km,\km)\in \set{(2,1,1),(3,1,0),(3,1,1)}$, and use Lemma~\ref{lm:existence} to show the existence of $\alpha$ which satisfies  \eqref{eq:condition1}. It is worth noting that when we apply Lemma~\ref{lm:existence}, the collection of polynomials under consideration  is not necessarily the set $\cF_b^{\kp.\km}$. We first look at the case of $b=2$ and $(\kp,\km)=(1,1)$.

\begin{theorem}\label{thm:b=2-kp=1-km=1}
For all sufficiently large prime powers $q$ such that $q\equiv 7\pmod{12}$, there is a perfect lattice code of $\Z^n$ with $n=(q-1)/6$, which can correct a single cyclic $2$-burst of $(1,1)$-limited-magnitude errors.
\end{theorem}

\begin{IEEEproof}
Recalling~\eqref{eq:defe}, \eqref{eq:defn}, and~\eqref{eq:deff}, in this case we have $e=6$, $q\equiv 1\pmod{6}$, and
\[\cF_2^{1,1}=\set*{1, 1+x^6, 1-x^6, -1, -1+x^6, -1-x^6}. \]
We label the polynomials in $\cF_2^{1,1}$   as $f_0,f_1,\ldots, f_5$, and then \eqref{eq:condition1} becomes
\[ \set{ \ord_\alpha(f_i(\alpha)) \ppmod{6} ; 0\leq i\leq 5 } = \set{0,1,\ldots,5}.        \]
Since $q\equiv 1 \pmod{6}$, for any primitive $\alpha$ we have
\[\ord_\alpha(-1) = (q-1)/2 \equiv 0 \pmod{3}.\]
Note that
\begin{align*}
\ord_{\alpha}(1)&=0, \\
\ord_{\alpha}(-1+\alpha^6)&\equiv
\ord_{\alpha}(-1)+\ord_\alpha(1-\alpha^6) \pmod{6},\\
\ord_\alpha(-1-\alpha^6)&\equiv \ord_{\alpha}(-1)+\ord_\alpha(1+\alpha^6)\pmod{6}.
\end{align*}
  Hence, in order to ensure \eqref{eq:condition1}, it suffices to require that   $q\equiv 7\pmod{12}$, i.e., $\ord_{\alpha}(-1)\equiv 3 \pmod{6}$, and
\begin{equation}\label{eq:condition2}
\set{ \ord_\alpha(1+\alpha^6) \ppmod{3}, \  \ord_\alpha(1-\alpha^6) \ppmod{3}  } = \set{1,2}.
\end{equation}

We shall use Lemma~\ref{lm:existence} to show the existence of $\alpha$ which satisfies \eqref{eq:condition2}. Then according to the discussion above and Proposition~\ref{prop:fieldcon}, the additive group of $\F_q$ can be split by $\cE^\circ(n,2,1,1)$ with $\vs_\alpha$, and so, the perfect $2$-burst-correcting code exists.

Consider the collection of polynomials $\cF=\set*{1+x^6, 1-x^6}$. Let $\ell_1=1,\ell_2=2$, and $h=3$. Let $\Theta$ be defined as in \eqref{eq:deftheta} for $\cF$.
For each $(i_1,i_2) \in \set*{0,1,2}^2 \setminus \set*{(0,0)}$, let
\[f_{i_1,i_2}(x) \eqdef (1+x^6)^{\frac{(q-1)i_1}{3}} (1-x^6)^{\frac{(q-1)i_2}{3}}.\]
It can be checked that the polynomials $f_{i_1,i_2}(x)$ satisfy the condition in Lemma~\ref{lm:existence}:
\begin{enumerate}
\item  If $i_2\not=0$, then $1-x$ is a factor of $f_{i_1,i_2}(x)$. Since $q\equiv 7 \pmod{12}$, we have that  $1-x \nmid 1+x^6$, and  $\gcd(1-x^6,-6x^5)=1$. Thus, in the canonical factorization of $f_{i_1,i_2}(x)$, the power of $1-x$ is $\frac{(q-1)i_2}{3}$, which is not a multiple of $q-1$. It follows that $f_{i_1,i_2}(x)$ cannot be written in the form $c (h(x))^{q-1}$.
\item If $i_2=0$, then $i_1\neq 0$. Since $\gcd(1+x^6,6x^5)=1$, in the factorization of $1+x^6$, every irreducible factor has power $1$. Thus, $f_{i_1,0}(x)$ cannot be written in the form $c (h(x))^{q-1}$.
\end{enumerate}
Applying  Lemma~\ref{lm:existence}, we get
\[\abs*{ \sum_{\alpha \in \F_q^*} \Theta(\alpha) -\phi(q-1)  } \leq A d(q-1)\sqrt{q},\]
which implies that
\[ \sum_{\alpha \in \F_q^*} \Theta(\alpha) \geq \phi(q-1)  - A d(q-1)\sqrt{q}.\]
Note that $A$ is independent of $q$, and for any given small $\varepsilon >0$ we have  $\phi(q-1)>q^{1-\varepsilon}$ and $d(q-1)<q^{\varepsilon}$ for all sufficiently large $q$ (see \cite[Theorem 315 and Theorem 327]{HarWri08}). Hence,
$\sum_{\alpha \in \F_q^*} \Theta(\alpha) >0,$
and so, there is an $\alpha \in \F_q^*$ such that $\Theta(\alpha) >0$.
According to the definition of $\Theta$, this $\alpha$ is the desired element to satisfy~\eqref{eq:condition2}.
\end{IEEEproof}

Now, we turn to the case of $b=3$ and $(\kp,\km)=(1,0)$. This time, using~\eqref{eq:defe}, \eqref{eq:defn}, and~\eqref{eq:deff}, we have $e=4$, $q\equiv 1\pmod{4}$, and
\[\cF_3^{1,0}=\set*{1,1+x^4,1+x^8,1+x^4+x^8}. \]
The idea is the same as before. We use Lemma~\ref{lm:existence} to find a primitive $\alpha$ such that the logarithm of the evaluations of the polynomials in $\cF_3^{1,0}$ at $\alpha$, are different modulo $4$. However, here we need to consider two different collections of polynomials when applying Lemma~\ref{lm:existence}, depending on whether $q$ is divisible by $3$ or not.

\begin{theorem}
\label{thm:b=3-kp=1-km=0-c}
For all sufficiently large prime powers $q$ such that $q\equiv 1\pmod{4}$, there is a perfect lattice code of $\Z^n$ with $n=(q-1)/4$, which can correct a single cyclic $3$-burst of $(1,0)$-limited-magnitude errors.
\end{theorem}

\begin{IEEEproof}
If $q$ is not divisible by $3$, consider the collection of polynomials
\[\cF=\cF_3^{1,0}\setminus \set*{1}=\set*{1+x^4,1+x^8,1+x^4+x^8},\]
as $\ord_{\alpha}(1)=0$ for all primitive $\alpha$.  Let $h=e=4$, $\ell_1=1$, $\ell_2=2$, $\ell_3=3$, and let $\Theta$ be defined as in \eqref{eq:deftheta}. Consider
\[f_{i_1,i_2,i_3}(x)\eqdef (1+x^4)^{\frac{(q-1)i_1}{4}} (1+x^8)^{\frac{(q-1)i_2}{4}}(1+x^4+x^8)^{\frac{(q-1)i_3}{4}}, \]
where $(i_1,i_2,i_3) \in \set{0,1,2,3}^3 \setminus \set{(0,0,0)}$.

We verify that $f_{i_1,i_2,i_3}(x)$ cannot be written as $c\cdot (h(x))^{q-1}$ for any $c\in \F_q$ and $h(x)\in \F_q[x]$:
\begin{enumerate}
\item If $i_1\neq0$, let $p(x)$ be an irreducible factor of $1+x^4$ and $a$ be a root of $p(x)$ in its splitting field. Since $\gcd(1+x^4,4x^3)=1$, the power of $p(x)$ in the factorization of $1+x^4$ is $1$. Moreover, $p(x)$ does not divide $(1+x^8)(1+x^4+x^8)$ as $(1+a^8)(1+a^4+a^8)=2 \neq 0$. Hence, in the factorization of $f_{i_1,i_2,i_3}(x)$, the power of $p(x)$ is $\frac{(q-1)i_1}{4}$, which is not a multiple of $q-1$.
\item If $i_1=0$ and $i_2\neq0$, let $p(x)$ be an irreducible factor of $1+x^8$. Using the same argument as above, we can show that  in the factorization of $f_{0,i_2,i_3}(x)$, the power of $p(x)$ is $\frac{(q-1)i_2}{4}$, which is not a multiple of $q-1$.
\item If $i_1=i_2=0$ and $i_3\neq 0$, $f_{0,0,i_3}(x)= (1+x^4+x^8)^{\frac{(q-1)i_3}{4}}$.   Note that $2(1+x^4+x^8)=(1+2x^4)(x^4-1)+3(1+x^4)$. Since $q$ is not divisible by $3$ and $\gcd(1+x^4,1+2x^4)=1$, we have $\gcd(1+x^4+x^8,1+2x^4)=1$, and so, $\gcd(1+x^4+x^8, 4x^3+8x^7)=1$. Hence,  in the factorization of $f_{0,0,i_3}(x)$ every irreducible factor has power $\frac{(q-1)i_3}{4}$, which is not a multiple of $q-1$.
\end{enumerate}

Then according to Lemma~\ref{lm:existence}, when $q$ is sufficiently large, there is a primitive element $\alpha$ such that the logarithm of the evaluations of the polynomials in $\cF_3^{1,0}$ at $\alpha$, are distinct modulo $4$. The conclusion then follows from Proposition~\ref{prop:fieldcon} and Theorem~\ref{th:lattotile}.

If $q$ is divisible by 3, we have $(1+x^4+x^8)=(1-x^4)^2$. Then it suffices to find a primitive element $\alpha$ such that
\begin{equation}\label{eq:condition3}
\ord_\alpha(1+\alpha^4) \equiv 1 \ppmod{4}, \quad \ord_\alpha(1+\alpha^8) \equiv 3 \ppmod{4}, \quad \ord_\alpha(1-\alpha^4) \equiv 1 \ppmod{4}.
\end{equation}
Let
\[g_{i_1,i_2,i_3}(x)\eqdef (1+x^4)^{\frac{(q-1)i_1}{4}} (1+x^8)^{\frac{(q-1)i_2}{4}} (1-x^4)^{\frac{(q-1)i_3}{4}},\]
where $(i_1,i_2,i_3) \in \set{0,1,2,3}^3 \setminus \set{(0,0,0)}$. If $i_3=0$, then  $g_{i_1,i_2,0}(x)=f_{i_1,i_2,0}(x)$, and so, it cannot be written as $c\cdot (h(x))^{q-1}$. If $i_3\neq 0$, since  $1-x \nmid (1+x^4)(1+x^8)$ and $1-x^4=(1-x)(1+x+x^2+x^3)$, in the factorization of $g_{i_1,i_2,i_3}(x)$ the factor $1-x$ has power $\frac{(q-1)i_3}{4}$, which is not a multiple of $q-1$. Hence, we can apply Lemma~\ref{lm:existence} with $\cF=\set{1+x^4,1+x^8,1-x^4}$ to show the existence of $\alpha$ such that \eqref{eq:condition3} holds, when $q$ is sufficiently large, which completes our proof.
\end{IEEEproof}

For the case of  $b=3$ and $\kp=\km=1$, we have  $e=18$ and $q\equiv 1 \pmod{18}$. Since $\ord_\alpha(-1)={(q-1)/2}$ for any primitive element $\alpha$, we shall assume $q\equiv 19\ppmod{36}$ such that $\ord_\alpha(-1) \not\equiv \ord_\alpha(1) \pmod{18}$.

\begin{theorem}
\label{thm:b=3-kp=1-km=1-c}
For all sufficiently large prime powers $q$ such that $q\equiv 19\pmod{36}$, there is a perfect lattice code of $\Z^n$ with $n=(q-1)/18$, which can correct a single cyclic $3$-burst of $(1,1)$-limited-magnitude errors.
\end{theorem}

\begin{IEEEproof}
The proof repeats the same steps taken in the previous two proofs. We therefore briefly sketch its outline. We have $e=18$ and let
\begin{align*}
f_1(x)&=1+x^e, &  f_2(x)&=1-x^e, &  f_3(x)&=1+x^{2e}, & f_4(x)&=1-x^{2e}, \\
 f_5(x)&=1+x^e+x^{2e}, & f_6(x)&=1-x^e+x^{2e},& f_7(x)&=1+x^e-x^{2e}, & f_8(x)&=1-x^e-x^{2e}.
\end{align*}
Since $\ord_\alpha(-1)\equiv 9 \pmod{18}$ and  $f_4(x)=1-x^{2e}=f_1(x)f_2(x)$, if we can find a primitive $\alpha$ such that
\begin{align*}
\ord_\alpha(f_1(\alpha)) &\equiv 1 \pmod{9},  & \ord_\alpha(f_2(\alpha)) &\equiv 2 \pmod{9},  \\
\ord_\alpha(f_3(\alpha)) &\equiv 6 \pmod{9}, & \ord_\alpha(f_5(\alpha)) &\equiv 5 \pmod{9}, \\
\ord_\alpha(f_6(\alpha)) &\equiv 7 \pmod{9},  & \ord_\alpha(f_7(\alpha)) &\equiv 4 \pmod{9},  \\
\ord_\alpha(f_8(\alpha)) &\equiv 8 \pmod{9},
\end{align*}
then  \eqref{eq:condition1} holds.

We set $h=9$. If $q$ is not divisible by $5$, it is verifiable that the set  of polynomials $\set{f_i ; 1\leq i\leq 8, i\neq 4}$ satisfies the condition of Lemma~\ref{lm:existence}. Thus, when $q$ is large enough, such an $\alpha$ exists.

If $q$ is divisible by $5$, then
\begin{align*}
f_3(x)&=1+x^{2e}=(x^e+2)(x^e-2),\\
f_7(x)&=1+x^e-x^{2e}=-(x^e+2)^2,\\
f_8(x)&=1-x^e+x^2e=-(x^e-2)^2. \end{align*}
Let $f_9(x)=x^e+2$ and $f_{10}(x)=x^e-2$.
Then the set of polynomials  $\set{f_1, f_2, f_5, f_6, f_9, f_{10}}$ satisfies the condition of Lemma~\ref{lm:existence}. Therefore, when $q$ is large enough, there is a primitive element $\alpha$ such that
\begin{align*}
\ord_\alpha(f_1(\alpha)) &\equiv 1 \pmod{9},  &
\ord_\alpha(f_2(\alpha)) &\equiv 2 \pmod{9}, \\
\ord_\alpha(f_5(\alpha)) &\equiv 5 \pmod{9}, &
\ord_\alpha(f_6(\alpha)) &\equiv 7 \pmod{9}, \\
\ord_\alpha(f_9(\alpha)) &\equiv 2 \pmod{9}, &
\ord_\alpha(f_{10}(\alpha)) &\equiv 4 \pmod{9}.
\end{align*}
It then follows that
\begin{align*}
\ord_\alpha(f_3(\alpha)) &\equiv 6 \pmod{9},  &
\ord_\alpha(f_4(\alpha)) &\equiv 3 \pmod{9}, \\
\ord_\alpha(f_7(\alpha)) &\equiv 4 \pmod{9}, &
\ord_\alpha(f_8(\alpha)) &\equiv 8 \pmod{9}.
\end{align*}
Hence, $\alpha$ is the desired primitive element.
\end{IEEEproof}

\subsection{Modification of the Constructions}

Theorem~\ref{thm:b=2-kp=1-km=1} shows the existence of lattice tilings of $\cE^{\circ}(n,2,1,1)$ when $q\equiv 7\pmod{12}$, whereas the necessary condition on $q$ is only $q \equiv 1\pmod{6}$. Thus, the existence of such tilings when $q\equiv 1 \pmod{12}$ remains undecided. In the following, we solve half of the remaining cases. We assume that $q=12m+1$ with $m$ odd, and show that a different splitting sequence provides a tiling. The following proposition is the equivalent of Proposition~\ref{prop:fieldcon}.

\begin{proposition}
\label{prop:fieldcon-m}
Assume $n\geq 3$, $q=12m+1$, $m$ odd, and define
\begin{align*}
\vr_{\alpha} &\eqdef (1, \alpha^3, \alpha^{12}, \alpha^{15}, \ldots, \alpha^{12(m-1)}, \alpha^{12(m-1)+3}), \\
\cF &\eqdef \set{\pm1,\pm x^3,\pm( 1+x^3), \pm(1-x^3), \pm(x^3+x^{12}), \pm(x^3-x^{12}) }.
\end{align*}
Let $\alpha$ be a primitive element of $\F_q^*$, and assume $f(\alpha)\neq 0$ for all $f(x)\in\cF$. If
\begin{equation}\label{eq:condition-m}
\set*{\ord_\alpha(f(\alpha)) \ppmod{12} ; f(x) \in \cF  } = \set*{0,1,2,\ldots,11},
\end{equation}
then $\cE^{\circ}(n,2,1,1)$ splits $G$ (the additive group of $\F_q$) with the splitting sequence $\vr_\alpha$.
\end{proposition}

\begin{IEEEproof}
For each  pair $\vc=(c_0,c_1)\in [-1,1]^2$ with $c_0\neq 0$, let
\begin{align*}
\cA_\vc & \eqdef \set*{\ve=(e_0,e_1,\ldots,e_{n-1}) \in \cE^{\circ}(n,2,1,1); \textup{ there is an \emph{even} integer $i$ such that }   \ve[i,i+1]=\vc }, \\
\cB_\vc &\eqdef \set*{\ve=(e_0,e_1,\ldots,e_{n-1}) \in \cE^{\circ}(n,2,1,1); \textup{ there is an \emph{odd} integer $i$ such that }   \ve[i,i+1]=\vc },
\end{align*}
where in both cases, $i+1$ is taken modulo $n$.
Then
\begin{align*}
\set*{\ve \cdot \vr_\alpha ; \ve \in \cA_{\vc} }   &= \set*{ \alpha^{12\ell} (c_0+c_1 \alpha^3); \ell \in [0,m-1]},\\
\set*{\ve \cdot \vr_\alpha ; \ve \in \cB_{\vc} }   &= \set*{ \alpha^{12\ell} (c_0\alpha^3+c_1 \alpha^{12}); \ell \in [0,m-1]}.
\end{align*}

Since \eqref{eq:condition-m} holds, we have that
\[\set*{\ve \cdot \vr_\alpha ; \ve \in \cE^{\circ}(n,2,1,1)} = \set{0} \cup \parenv*{ \bigcup_{\substack{\vc \in [-1,1]^{2} \\ c_0\neq 0}}\parenv*{ \set*{\ve \cdot \vr_\alpha ; \ve \in \cA_{\vc} } \cup  \set*{\ve \cdot \vr_\alpha ; \ve \in \cB_{\vc} } }  }=\F_q.\]
Hence $\cE^{\circ}(n,2,1,1)$ splits $G$ with $\vr_\alpha$.
\end{IEEEproof}

\begin{theorem}\label{thm:b=2-kp=1-km=1-2}
For all sufficiently large prime powers $q$ such that $q\equiv 13\pmod{24}$, there is a perfect lattice code of $\Z^n$ with $n=(q-1)/6$, which can correct a single  cyclic $2$-burst  of $(1,1)$-limited-magnitude errors.
\end{theorem}

\begin{IEEEproof}
Since $q\equiv 13 \pmod{24}$, $\ord_{\alpha}(-1)=(q-1)/2 \equiv 6 \pmod{12}$. Thus if the logarithms of $1,\alpha^3, 1+\alpha^3, 1-\alpha^3, \alpha^3+\alpha^{12}, \alpha^3-\alpha^{12}$ are distinct modulo $6$, then~\eqref{eq:condition-m} holds. To find such a primitive $\alpha$, we let $h=6$ and consider the following set of polynomials
\[\cF'\eqdef \set*{1+x^3, 1-x^3, 1+x^3+x^6, 1-x^3+x^6}.\]
It is verifiable that these polynomials satisfy the condition in Lemma~\ref{lm:existence}. Hence, if $q$ is large enough, there is a primitive $\alpha$ such that
\begin{align*}
   \ord_{\alpha}(1+\alpha^3)&\equiv 1 \ppmod{6}, &\ord_{\alpha}(1-\alpha^3)&\equiv 2 \ppmod{6}, \\
   \ord_{\alpha}(1-\alpha^3+\alpha^6) &\equiv 0 \ppmod{6}, & \ord_{\alpha}(1+\alpha^3+\alpha^6)&\equiv 0 \ppmod{6}.
\end{align*}
Then it follows that
\begin{align*}
\ord_{\alpha}(\alpha^3+\alpha^{12}) & \equiv 3+\ord_{\alpha}(1+\alpha^3)+\ord_{\alpha}(1-\alpha^3+\alpha^6)\equiv 4 \ppmod{6},\\
\ord_{\alpha}(\alpha^3-\alpha^{12}) &\equiv 3+\ord_{\alpha}(1-\alpha^3)+\ord_{\alpha}(1+\alpha^3+\alpha^6)\equiv 5 \ppmod{6}.
\end{align*}
Noting that $\ord_{\alpha}(1)=0$ and $\ord_{\alpha}(\alpha^3)=3$, we have completed our proof.
\end{IEEEproof}

\section{Discussion}
\label{sec:comments}

In this paper we constructed perfect lattice codes that are capable of correcting a single burst of limited-magnitude errors. Our constructions span both the case of cyclic burst errors, as well as non-cyclic bursts. The parameters of the various constructions are summarized in Table~\ref{tab:summary}. We note that the first row in this table is obtained by a standard argument that converts a code over $\F_p$, $p$ a prime, to a lattice code.

\begin{table*}
\center
  \caption{Summary of perfect-code constructions ($q$ is a prime power)}
  \label{tab:summary}
  {\renewcommand{\arraystretch}{1.2}
  \begin{tabular}{cccccll}
    \hline\hline
      $b$ & $\kp$ & $\km$ & $n$ & Cyclic & Source & Comments \\
    \hline
      $2$ & $1$ & $0$ & $n=2^{r}$ & N & \cite{Etz01b} & $r\geq 4$\\
      $2$ & $1$ & $0$ & $n \geq 2$ & N & Theorem~\ref{thm:b=2kp=1km=0-nc} & \\
      $2$ & $1$ & $0$ & $4\leq n\equiv 1,4 \ppmod{6}$ & Y & Theorem~\ref{thm:b=2kp=1km=0-c2} & \\
      $2$ & $1$ & $0$ & $n=\frac{q-1}{2}$ & Y & Theorem~\ref{thm:b=2kp=1km=0-c1} & $q\geq 7$ odd \\
      $2$ & $1$ & $1$ & $n=\frac{q-1}{6}$ & Y & Theorem~\ref{thm:b=2-kp=1-km=1} & $q\equiv 7 \ppmod{12}$ sufficiently large\\
      $2$ & $1$ & $1$ & $n=\frac{q-1}{6}$ & Y & Theorem~\ref{thm:b=2-kp=1-km=1-2} & $q\equiv 13 \ppmod{24}$ sufficiently large\\
      $3$ & $1$ & $0$ & $n=\frac{q-1}{4}$ & Y & Theorem~\ref{thm:b=3-kp=1-km=0-c} & $q\equiv 1 \ppmod{4}$ sufficiently large\\
      $3$ & $1$ & $1$ & $n=\frac{q-1}{18}$ & Y & Theorem~\ref{thm:b=3-kp=1-km=1-c} & $q\equiv 19 \ppmod{36}$ sufficiently large\\
    \hline\hline
  \end{tabular}
  }
\end{table*}

The approach in Section~\ref{sec:confield} was inspired by~\cite{AbdMcEOdlTil86}. This is in particular interesting, since~\cite{AbdMcEOdlTil86} did not study perfect codes. Similar to~\cite{AbdMcEOdlTil86}, our constructions in Section~\ref{sec:confield} call for finding a primitive element of $\F_q$ with certain properties. We note that a simple brute-force search can easily find such an element (if it exists) in time polynomial in $q$, which is also polynomial in $n$ as $n=\Theta(q)$ in all of our constructions. The number-theoretic conditions required by our constructions seem to make it difficult to give an existence guarantee stronger than ``sufficiently large $q$''. We ran a computer search, whose results are summarized in Table~\ref{tab:compsearch}. The table count the number of good prime powers (i.e., those that admit a primitive $\alpha$ with the required properties), the number of bad prime powers, and the list of bad prime powers.

\begin{table*}
\center
  \caption{Values of $q$, from $e(2b-1)+1$ up to $1000$, that do not admit the required primitive element}
  \label{tab:compsearch}
  {\renewcommand{\arraystretch}{1.2}
  \begin{tabular}{cccl}
    \hline\hline
      & \#Good & \#Bad & Bad Prime Powers \\
    \hline
      Theorem~\ref{thm:b=2-kp=1-km=1} & 41 & 3 &
      $19, 43, 127$ \\
      Theorem~\ref{thm:b=2-kp=1-km=1-2} & 6 & 15 &
      $37, 61, 109, 157, 181, 229, 277, 349, 373, 397, 421, 613, 661,  733, 829$ \\
      Theorem~\ref{thm:b=3-kp=1-km=0-c} & 76 & 14 &
      $25, 37, 49, 61, 97, 101, 121, 157, 169, 289, 361, 449, 601, 729$ \\
      Theorem~\ref{thm:b=3-kp=1-km=1-c} & 2 & 13 &
      $199, 271, 307, 343, 379, 487, 523, 631, 739, 811, 883, 919, 991$ \\
    \hline\hline
  \end{tabular}
  }
\end{table*}

We would also like to comment on the prospect of extending our constructions, both for longer bursts, as well as for errors of larger magnitude.

\subsection{Longer Bursts}

In Section~\ref{sec:confield} we presented a construction based on finite fields and used it to prove  a few existence results for lattice tiling of $\cE^{\circ}(n,t,\kp,\km)$ with $b\leq 3$ and $(\kp,\km)\in \set{(1,0),(1,1)}$. This approach may also work for the  cases  $b>3$. However,  it would involve choosing a large number of factors of the polynomials in $\cF_{b}^{\km,\km}$, checking whether they satisfy the condition in Lemma~\ref{lm:existence}, and assigning each of them an integer such that \eqref{eq:condition1} holds. Thus, a closed-form solution to all the cases $b>3$ still remains unsolved. We note that a similar problem was considered in~\cite{AbdMcEOdlTil86} for polynomials that satisfy the AES conditions,  and it was solved by  showing that  it suffices to consider only irreducible polynomials and assign all of them the same integer zero~\cite[Theorem 3]{AbdMcEOdlTil86}. Whether a similar solution exists here is still unknown.

\subsection{Larger Error Magnitudes}

In this paper, we studied only the case $\kp=1$. For $\kp\geq 2$,  finding a lattice tiling becomes more difficult. If one wants to use the construction in Section~\ref{sec:confield} to handle the case of $b=2$ and $(\kp,\km)=(2,0)$, a primitive element $\alpha$ satisfying the following condition is required:
\begin{equation} \label{eq:condition4}
\set{ \ord_\alpha(f_i(\alpha)) \ppmod{6} ; 1\leq i\leq 5 } = \set{1,2,3,4,5},
\end{equation}
where
\begin{align*}
f_1(x)&=1+x^6, & f_2(x)&=1+2x^6, & f_3(x)&=2,\\
f_4(x)&=2+x^6, & f_5(x)&=2+2x^6.
\end{align*}
Note that unlike $\ord_{\alpha}(1)=0$ and $\ord_{\alpha}(-1)=\frac{q-1}{2}$, the value of $\ord_{\alpha}(2)$  modulo $6$ depends on the choice of $\alpha$. To complicate things further, $f_3(x)=2$ does not satisfy the condition in Lemma~\ref{lm:existence}. Thus, we cannot use it, as is, to find the desired $\alpha$. A computer search up to $1000$ shows that the following field sizes, $q$,
\begin{align*}
& 19, 79, 103, 163, 181, 199, 229, 349, 373, 397, 421, 487, 499, 541, 613, 619, 631, 643,   691, 709, 733, 739, 751, 769, \\ & 787, 823, 853, 859, 907, 967, 997
\end{align*}
admit a primitive $\alpha$ that satisfies~\eqref{eq:condition4}.

We also ran a computer search for splittings by $\cE^\circ(n,2,2,0)$ and $\cE(n,2,2,0)$. For $n\in \set{3,4}$, existence results are listed in Table~\ref{tab:b=2-kp=2-km=0}. Interestingly, for each $5\leq n \leq 11$, every Abelian group $G$ of order $6n+1$  cannot be split by $\cE^{\circ}(n,2,2,0)$, and every Abelian group of order $6n-3$ cannot be split by $\cE(n,2,2,0)$. In contrast, for the case of $b=2$ and $(\kp,\km)=(1,1)$, Table~\ref{tab:b=2-kp=1-km=1-c} shows that $\Z_{6n+1}$ can be split by  $\cE^{\circ}(n,2,1,1)$  for each $n \in [4,14] \setminus \set{7}$, and Table~\ref{tab:b=2-kp=1-km=1-nc} shows that $\Z_{6n-3}$ can be split by  $\cE(n,2,1,1)$ for each $n \in [3, 14]$. Thus, it would be interesting to derive  some constraints on the values of $n$ for the existence of lattice tilings of  $\cE(n,2,2,0)$ and $\cE^{\circ}(n,2,2,0)$.

\begin{table*}
\center
  \caption{Splitting of $G$ by  $\cE^{\circ}(n,2,2,0)$ or $\cE(n,2,2,0)$ }
  \label{tab:b=2-kp=2-km=0}
  {\renewcommand{\arraystretch}{1.2}
  \begin{tabular}{cccl}
    \hline\hline
      $n$ & $G$  & The shape & Splitting sequence \\
    \hline
      $3$ & $\Z_{19}$ & $\cE^{\circ}(n,2,2,0)$ & $\vs=(1,7,11)$ \\
      $4$ & $\Z_{25}$ & $\cE^{\circ}(n,2,2,0)$ & $\vs=(1,5,4,20)$ \\
      \hline
      $3$ & $\Z_{15}$ & $\cE(n,2,2,0)$ & $\vs=(1,5,4)$ \\
      $4$ & $\Z_{21}$ & $\cE(n,2,2,0)$ & $\vs=(1,5,20,18)$ \\
    \hline\hline
  \end{tabular}
  }
\end{table*}

\begin{table*}
\center
  \caption{Splitting of $\Z_{6n+1}$ by  $\cE^{\circ}(n,2,1,1)$}
  \label{tab:b=2-kp=1-km=1-c}
  {\renewcommand{\arraystretch}{1.2}
  \begin{tabular}{cl}
    \hline\hline
     $n$    & Splitting sequence \\
    \hline
     $4$    & $\vs=(1, 5, 2, 10)$\\
     $5$    & $\vs=(1, 4, 15, 2, 8)$\\
     $6$    & $\vs=(1, 8, 10, 6, 11, 14)$\\
     $8$    & $\vs=(1, 4, 21, 9, 2, 18, 8, 14)$\\
     $9$    & $\vs=(1, 3, 12, 25, 6, 20, 27, 17, 22)$\\
     $10$   & $\vs=(1, 3, 11, 24, 9, 25, 30, 12, 29, 22)$\\
     $11$   & $\vs=(1, 3, 9, 27, 14, 25, 8, 24, 5, 15, 22)$ \\
     $12$   & $\vs=(1, 3, 8, 27, 33, 12, 30, 20, 29, 7, 32, 15)$ \\
     $13$   & $\vs=(1, 3, 8, 14, 37, 17, 10, 26, 38, 9, 39, 21, 34)$\\
     $14$   & $\vs=(1, 3, 8, 14, 31, 7, 41, 9, 21, 39, 10, 23, 42, 27)$\\
    \hline\hline
  \end{tabular}
  }
\end{table*}

\begin{table*}
\center
  \caption{Splitting of $\Z_{6n-3}$ by  $\cE(n,2,1,1)$}
  \label{tab:b=2-kp=1-km=1-nc}
  {\renewcommand{\arraystretch}{1.2}
  \begin{tabular}{cl}
    \hline\hline
     $n$    & Splitting sequence \\
    \hline
     $3$    & $\vs=(1,5,2)$\\
     $4$    & $\vs=(1, 4, 10, 2)$\\
     $5$    & $\vs=(1, 4, 10, 2, 9)$\\
     $6$    & $\vs=(1, 14, 10, 2, 5, 11)$ \\
     $7$    & $\vs=(1, 3, 12, 19, 6, 16, 5)$\\
     $8$    & $\vs=(1, 3, 12, 20, 14, 21, 5, 22)$\\
     $9$    & $\vs=(1, 3, 9, 16, 5, 24, 10, 23, 8)$\\
     $10$   & $\vs=(1, 3, 8, 25, 13, 28, 6, 20, 27, 9)$\\
     $11$   & $\vs=(1, 3, 8, 29, 7, 25, 15, 28, 16, 30, 24)$\\
     $12$   & $\vs=(1, 3, 8, 17, 32, 13, 29, 7, 28, 18, 12, 26)$\\
     $13$   & $\vs=(1, 3, 8, 14, 32, 19, 31, 16, 26, 9, 30, 7, 27)$\\
     $14$   & $\vs=(1, 3, 8, 14, 30, 13, 40, 21, 12, 35, 10, 39, 24, 31)$\\
    \hline\hline
  \end{tabular}
  }
\end{table*}

\bibliographystyle{IEEEtrans}
\bibliography{allbib}

\end{document}